\documentclass[conference,11pt]{IEEEtran}

\usepackage[utf8]{inputenc}
\usepackage[T1]{fontenc}
\usepackage{amsmath,amssymb}
\usepackage{graphicx}
\usepackage{booktabs}
\usepackage{multirow}
\usepackage{array}
\usepackage{hyperref}
\usepackage{xcolor}
\usepackage{listings}
\usepackage{tikz}
\usetikzlibrary{shapes.geometric, arrows.meta, positioning, fit, calc, backgrounds, decorations.pathreplacing}
\usepackage{pgfplots}
\pgfplotsset{compat=1.18}
\usepackage{subcaption}
\usepackage{tabularx}
\usepackage{float}
\usepackage{algorithm}
\usepackage{algpseudocode}
\usepackage{balance}
\usepackage{enumitem}
\usepackage[font=small]{caption}
\usepackage{adjustbox}

\definecolor{phaseA}{HTML}{4472C4}
\definecolor{phaseB}{HTML}{ED7D31}
\definecolor{phaseC}{HTML}{A5A5A5}
\definecolor{phaseD}{HTML}{FFC000}
\definecolor{phaseE}{HTML}{5B9BD5}
\definecolor{phaseF}{HTML}{70AD47}

\hypersetup{
  colorlinks=true,
  linkcolor=blue!70!black,
  citecolor=blue!70!black,
  urlcolor=blue!60!black,
  pdftitle={Engineering Lessons from Authorized YouTube-to-Blockchain Content Replication at Scale},
  pdfauthor={Muhammad Zeeshan Akram},
  pdfsubject={Cross-platform content replication, web scraping at scale, anti-detection, decentralized video, blockchain storage},
  pdfkeywords={YouTube, web scraping, anti-detection, proxy rotation, rate limiting evasion, blockchain, decentralized storage, content migration, platform independence, fault tolerance, OAuth, bot detection},
}

\begin{document}

\title{Engineering Lessons from Authorized\\YouTube-to-Blockchain Content Replication at Scale}

\author{
\IEEEauthorblockN{Muhammad Zeeshan Akram}
\IEEEauthorblockA{University of Louisville \\
Formerly at JSgenesis / Joystream DAO \\
\textit{muhammadzeeshan.akram@louisville.edu}}
}

\maketitle

\begin{abstract}
We present an experience report on \textbf{YouTube-Synch}\footnote{Source code: \url{https://github.com/Joystream/youtube-synch}}, a production system that replicates content from 10{,}000+ \emph{creator-authorized} YouTube channels to a blockchain-based decentralized platform. Although replication is authorized, the system must still defeat YouTube's anti-automation defenses---API quota restrictions (10{,}000 units/day), IP-based rate limiting, behavioral bot detection, and OAuth token lifecycle policies---which do not distinguish authorized bulk access from abuse.

Our central observation is a \emph{coupled-defense} phenomenon: these protection layers are not independent, so circumventing one (e.g., API quotas) silently activates another (e.g., OAuth token expiration), producing cascading, delayed failures. We ground this in three production incidents with concrete impact---28 duplicate on-chain objects from a database throughput failure, 10{,}000+ channels lost to a single OAuth mass-expiration, and 719 daily errors from queue pollution---observed over 15 releases and 3.5 years of operation.

We further argue, from the system's own concurrency and rate parameters alone, that detection-driven anti-bot measures impose a download-bound throughput ceiling on the order of $10^3$ videos/day per instance---at or below the steady-state demand of 10{,}000 channels. This analysis shows why priority-based triage and horizontal scaling are structural necessities rather than optimizations, and explains the survivability-over-throughput tradeoff that drove a 25$\times$ reduction in download concurrency. We detail the resulting architecture---a four-stage DAG pipeline, a Write-Ahead-Log fault-tolerance model with cross-system state reconciliation, and a trust-minimized ownership-verification protocol that eliminates OAuth---and distill design principles that generalize to extraction against other heavily-defended centralized platforms.
\end{abstract}

\begin{IEEEkeywords}
Web scraping at scale, anti-detection, rate limiting evasion, proxy rotation, bot detection, platform defense circumvention, decentralized video, blockchain content replication, YouTube API, cross-platform content migration, OAuth security, fault tolerance, distributed storage
\end{IEEEkeywords}

\section{Introduction}
\label{sec:intro}

YouTube, with over 800 million videos and 2.7 billion monthly active users~\cite{youtube-stats}, represents not merely a dominant video platform but a multi-billion dollar infrastructure fortress---one whose content protection mechanisms have been refined over nearly two decades to prevent precisely the kind of large-scale automated content extraction that decentralized platforms need to bootstrap their ecosystems. The growing deplatforming phenomenon~\cite{deplatforming} has increased creator interest in platform independence, yet without a critical mass of quality content, decentralized platforms cannot attract viewers; without viewers, creators have no incentive to migrate.

\textbf{YouTube-Synch} was designed to breach this fortress---with creator authorization. Rather than requiring creators to manually re-upload their content, the system continuously monitors their YouTube channels, downloads new videos, creates corresponding on-chain representations on the Joystream blockchain~\cite{joystream}, and uploads assets to distributed storage nodes. This \emph{sync-not-migrate} approach preserves the creator's existing YouTube audience while simultaneously building their decentralized presence.

The system operates as the technical backbone of the \textbf{YouTube Partner Program (YPP)}, which incentivizes YouTube creators to authorize content replication through a tiered reward structure. Since its initial release in August 2022, YouTube-Synch has grown from a prototype processing a handful of channels to a production system managing over 10{,}000 enrolled channels, evolving through 15 release versions, 144 merged pull requests, and 203 tracked issues across 3.5 years of active development.

The core narrative of this system's evolution is a sustained cat-and-mouse game with YouTube's infrastructure. Each layer of YouTube's defense---API quotas, IP blocking, bot detection, OAuth restrictions---demanded a distinct architectural countermeasure, and the interaction between these defenses produced cascading failures that reshaped the entire system:

\begin{itemize}[nosep]
  \item \textbf{API quota strangulation}: YouTube Data API v3 imposes a hard daily quota of 10{,}000 units per project. At 10{,}000+ enrolled channels, the system cannot afford even one API call per channel per day---forcing a complete migration away from the official API.
  \item \textbf{IP-based rate limiting and bot detection}: After decoupling from the API, direct \texttt{yt-dlp} scraping at scale triggers YouTube's behavioral analysis, resulting in ``Sign in to confirm you're not a bot'' blocks. This forced three generations of proxy infrastructure (direct $\rightarrow$ Chisel tunnel $\rightarrow$ proxychains4 pool) and a 25$\times$ reduction in download concurrency (50 $\rightarrow$ 2).
  \item \textbf{OAuth token lifecycle weaponization}: Google's 6-month unused token expiration policy, combined with the API migration that rendered OAuth tokens unused, produced a mass-expiration event that opted out 10{,}000+ channels in a single polling cycle---forcing the complete elimination of Google OAuth in favor of a trust-minimized video-based verification protocol.
  \item \textbf{Blockchain throughput constraints}: Joystream's ${\sim}6$-second block time limits on-chain video creation throughput, necessitating batch transaction strategies with 10 extrinsics per batch.
  \item \textbf{Distributed storage eventual consistency}: The storage layer's eventual consistency model causes upload failures for recently-created on-chain objects, demanding retry-based resilience patterns with exponential backoff.
\end{itemize}

This paper contributes:
\begin{enumerate}[nosep]
  \item \textbf{The coupled-defense phenomenon}: evidence that a platform's defense-in-depth layers interact, so circumventing one activates another with delayed, cascading effects---grounded in three production incidents (Section~\ref{sec:challenges}).
  \item \textbf{An analytical scaling argument}: a throughput ceiling derived from the system's own parameters showing that detection limits, not blockchain throughput, force priority triage and horizontal scaling (Section~\ref{sec:ceiling}).
  \item \textbf{The survivability-over-throughput design principle} for sustained extraction against adaptive bot detection (Section~\ref{sec:anti-detection}).
  \item \textbf{A trust-minimized ownership-verification protocol} that eliminates OAuth, with a zkSNARK-based attribution design for the untrusted-operator setting (Section~\ref{sec:trust}).
  \item \textbf{A cross-system fault-tolerance model} reconciling an immutable blockchain target with a defended source via a Write-Ahead Log (Section~\ref{sec:fault-tolerance}).
  \item \textbf{A 3.5-year longitudinal experience report} with design principles that generalize to other defended platforms (Sections~\ref{sec:evolution},~\ref{sec:discussion}).
\end{enumerate}

\section{Background and Related Work}
\label{sec:background}

\subsection{Joystream Blockchain}

Joystream~\cite{joystream} is a Substrate-based~\cite{substrate} blockchain designed for decentralized video content governance and distribution. Its architecture comprises:

\begin{itemize}[nosep]
  \item \textbf{Content Directory}: An on-chain registry of channels, videos, and associated metadata, managed through Substrate extrinsics.
  \item \textbf{Colossus Storage Nodes}: Distributed storage infrastructure that accepts and serves video assets, conceptually similar to content-addressed storage systems like IPFS~\cite{benet2014ipfs} and Filecoin~\cite{filecoin}. Storage nodes are operated by elected storage providers.
  \item \textbf{Orion (Query Node)}: A GraphQL indexer that provides queryable access to processed on-chain state. The system migrated from the original Query Node to Orion in version 3.6.
  \item \textbf{DAO Governance}: A council-based governance system that manages program parameters including YPP reward budgets and tier structures.
\end{itemize}

\subsection{YouTube Data API v3}

The YouTube Data API v3 provides programmatic access to YouTube's content metadata. The API operates on a quota system where each project receives 10{,}000 units per day. Table~\ref{tab:api-costs} shows the cost structure that constrained the system's early architecture.

\begin{table}[h]
\centering
\caption{YouTube Data API v3 Quota Costs}
\label{tab:api-costs}
\begin{adjustbox}{max width=\columnwidth}
\begin{tabular}{@{}lcc@{}}
\toprule
\textbf{Operation} & \textbf{Cost (units)} & \textbf{Max calls/day} \\
\midrule
\texttt{channels.list}  & 1--5   & 2{,}000--10{,}000 \\
\texttt{search.list}    & 100    & 100 \\
\texttt{videos.list}    & 1--7   & 1{,}428--10{,}000 \\
\texttt{playlistItems.list} & 1--3 & 3{,}333--10{,}000 \\
OAuth token exchange     & 0      & Unlimited \\
\bottomrule
\end{tabular}
\end{adjustbox}
\end{table}

At 10{,}000 channels, a single \texttt{channels.list} call per day per channel consumes the entire daily quota, leaving zero budget for video discovery, signup flows, or error recovery.

\subsection{YouTube's Content Protection Infrastructure}
\label{sec:yt-defenses}

Understanding YouTube-Synch's architecture requires understanding the multi-layered defense infrastructure it must navigate. YouTube's protections are not a single wall but a defense-in-depth system where each layer interacts with the others, creating emergent barriers that are not apparent from examining any single layer in isolation:

\begin{enumerate}[nosep]
  \item \textbf{API Quota System}: The 10{,}000 unit/day quota is both a resource allocation mechanism and a rate limiter. The quota-per-operation cost structure (Table~\ref{tab:api-costs}) particularly penalizes search operations (100 units each), making channel discovery via the API prohibitively expensive. Importantly, quota exhaustion is \emph{silent}---requests return HTTP~403 without advance warning.

  \item \textbf{IP-Based Rate Limiting}: YouTube monitors request patterns at the IP level. Automated downloading tools such as \texttt{yt-dlp} must contend with progressive blocking: initial requests succeed, but sustained volume from a single IP triggers a ``Sign in to confirm you're not a bot'' challenge page within hours. The detection sensitivity is adaptive---higher volume triggers faster blocking.

  \item \textbf{Behavioral Fingerprinting}: Beyond simple rate counting, YouTube's bot detection analyzes request timing patterns, concurrent connection counts, and User-Agent consistency. Perfectly regular request intervals are a stronger bot signal than high volume with natural variance, making naive parallel downloading self-defeating.

  \item \textbf{OAuth Token Lifecycle}: Google's OAuth2 implementation includes several protective mechanisms: refresh tokens expire after 6 months of inactivity, access tokens expire after 1 hour, and projects in ``testing'' OAuth consent status have tokens that expire after 7 days. These create time bombs for any system that intermittently uses OAuth-based access.

  \item \textbf{Content Delivery Protections}: YouTube's video serving infrastructure employs signed URLs with time-limited validity, adaptive bitrate streaming that complicates direct download, and format negotiation that varies by geographic region and client profile.

  \item \textbf{Legal/TOS Framework}: YouTube's Terms of Service explicitly prohibit automated downloading except through the official API, creating a legal dimension that interacts with the technical defenses---circumventing IP blocks through proxies operates in a legally gray area that the creator-authorization model partially addresses.
\end{enumerate}

The critical insight driving YouTube-Synch's architectural evolution is that these layers are \emph{coupled}: circumventing the API quota by switching to direct scraping activates the IP blocking layer; circumventing IP blocking through proxies increases the surface area for behavioral fingerprinting; and the migration away from the API causes OAuth tokens to expire, activating the token lifecycle defense. Each countermeasure creates exposure to a different defense layer, requiring architectural responses that address the entire defense stack holistically.

\subsection{Related Work}

\textbf{LBRY/Odysee ytsync}~\cite{lbry} provided the earliest reference implementation for YouTube-to-decentralized content replication. Its architecture was studied during YouTube-Synch's design phase, but LBRY's approach differed fundamentally: it operated as a centralized service without creator authorization, raising IP and legal concerns that YouTube-Synch's opt-in model avoids.

\textbf{PeerTube}~\cite{peertube} implements federated video hosting using ActivityPub, supporting YouTube video import via \texttt{youtube-dl}. However, PeerTube's import is manual and one-shot---it does not provide continuous synchronization or channel-level automation.

\textbf{DTube}~\cite{dtube} integrates with IPFS/Skynet for decentralized video storage but relies on manual upload by creators rather than automated cross-platform replication.

The \textbf{youtube-dl} ecosystem~\cite{youtube-dl}, including the actively-maintained \texttt{yt-dlp} fork~\cite{yt-dlp}, provides the foundational video downloading capability that YouTube-Synch leverages. The evolution from direct YouTube API usage to \texttt{yt-dlp}-based scraping is central to the system's architectural story.

\section{System Architecture}
\label{sec:architecture}

\subsection{High-Level Overview}

\begin{figure*}[t]
\centering
\begin{tikzpicture}[
  node distance=0.6cm and 0.8cm,
  box/.style={draw, rounded corners=3pt, minimum width=2.2cm, minimum height=0.7cm, font=\footnotesize, align=center, fill=#1!15},
  box/.default=blue,
  svc/.style={draw, rounded corners=3pt, minimum width=4.5cm, minimum height=3cm, font=\footnotesize, dashed, thick, fill=#1!5},
  svc/.default=blue,
  infra/.style={draw, rounded corners=3pt, minimum width=2cm, minimum height=0.7cm, font=\footnotesize, fill=gray!15},
  arr/.style={-{Stealth[length=2.5mm]}, thick},
  darr/.style={-{Stealth[length=2.5mm]}, thick, dashed},
]

\node[box=red, minimum width=10cm, minimum height=0.7cm] (yt) at (0, 5) {\textbf{YouTube Platform} (Data API v3 / Operational API / yt-dlp)};

\node[svc=blue, minimum width=5cm, minimum height=3.5cm, label={[font=\footnotesize\bfseries]above:Sync Service (Docker)}] (sync) at (-3.5, 2) {};
\node[box=blue, minimum width=2.2cm] (poll) at (-3.5, 3.2) {YT Polling};
\node[box=blue, minimum width=2.2cm] (dl) at (-4.5, 2.2) {Download};
\node[box=blue, minimum width=2.2cm] (meta) at (-2.5, 2.2) {Metadata};
\node[box=blue, minimum width=2.2cm] (create) at (-4.5, 1.2) {Creation};
\node[box=blue, minimum width=2.2cm] (upload) at (-2.5, 1.2) {Upload};

\node[svc=green, minimum width=5cm, minimum height=3.5cm, label={[font=\footnotesize\bfseries]above:HTTP API Service (Docker)}] (api) at (3.5, 2) {};
\node[box=green, minimum width=2.2cm] (rest) at (3.5, 3.2) {NestJS REST};
\node[box=green, minimum width=2.2cm] (bull) at (2.5, 2.2) {BullMQ Board};
\node[box=green, minimum width=2.2cm] (swagger) at (4.5, 2.2) {Swagger Docs};
\node[box=green, minimum width=2.2cm] (signup) at (3.5, 1.2) {Signup / Auth};

\node[infra] (dynamo) at (-4.5, -0.8) {DynamoDB};
\node[infra] (redis) at (-1.5, -0.8) {Redis 7.2.1};
\node[infra] (chain) at (1.5, -0.8) {Joystream Chain};
\node[infra] (orion) at (4.5, -0.8) {Orion (GraphQL)};
\node[infra] (proxy) at (-4.5, -1.8) {SOCKS5 Proxies};
\node[infra] (elastic) at (-1.5, -1.8) {Elasticsearch};
\node[infra] (storage) at (1.5, -1.8) {Colossus Storage};
\node[infra] (hubspot) at (4.5, -1.8) {HubSpot CRM};

\draw[arr] (yt.south) -- ++(0,-0.3) -| (poll.north);
\draw[arr] (poll) -- (dl);
\draw[arr] (dl) -- (meta);
\draw[arr] (meta) -- (create);
\draw[arr] (create) -- (upload);
\draw[arr, dashed] (yt.south) -- ++(0,-0.3) -| (signup.north);

\draw[darr] (sync.south) -- (dynamo.north);
\draw[darr] (sync.south) -- (redis.north);
\draw[darr] (api.south) -- (dynamo.north);
\draw[darr] (api.south) -- (redis.north);
\draw[darr] (create.south) |- (chain.north);
\draw[darr] (upload.south) |- (storage.north);

\end{tikzpicture}
\caption{YouTube-Synch split-service architecture (v3.4+). The Sync Service handles content processing through a four-stage DAG pipeline, while the HTTP API Service manages creator onboarding, channel state, and operational dashboards. Both share DynamoDB for state and Redis for job queue coordination.}
\label{fig:architecture}
\end{figure*}
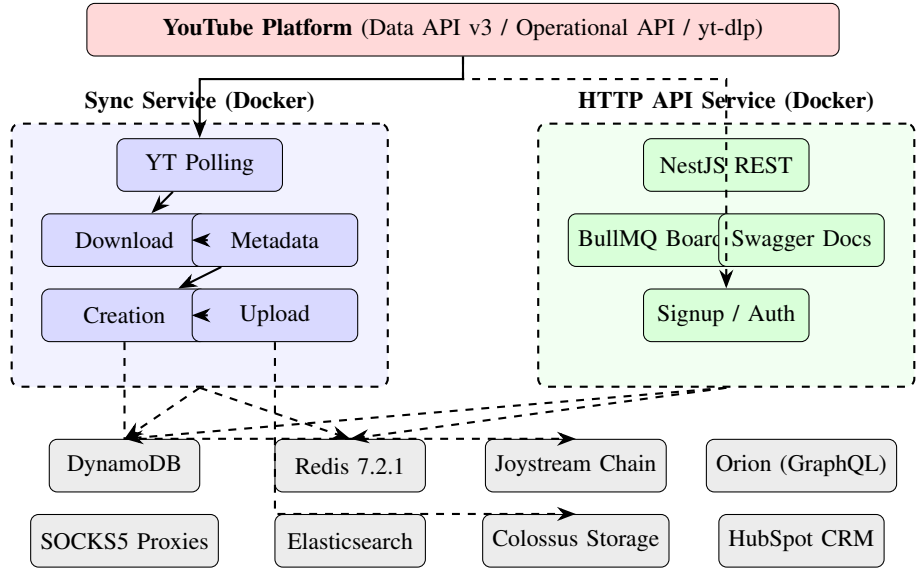

YouTube-Synch consists of two independently deployable services sharing a common data layer, as shown in Figure~\ref{fig:architecture}.

\subsection{Four-Stage Processing Pipeline}

The core of YouTube-Synch is a four-stage directed acyclic graph (DAG) processing pipeline implemented using BullMQ~\cite{bullmq} flow jobs. Each video traverses four queues, with the output of each stage triggering the next. Table~\ref{tab:pipeline-config} summarizes the production configuration.

\begin{table}[h]
\centering
\caption{Pipeline Stage Configuration (Production)}
\label{tab:pipeline-config}
\begin{adjustbox}{max width=\columnwidth}
\begin{tabular}{@{}llccl@{}}
\toprule
\textbf{Stage} & \textbf{Queue} & \textbf{Conc.} & \textbf{Timeout} & \textbf{Type} \\
\midrule
1. Download  & \texttt{DownloadQueue}  & 2   & 10{,}800\,s & concurrent \\
2. Metadata  & \texttt{MetadataQueue}  & 2   & 1{,}800\,s  & concurrent \\
3. Creation  & \texttt{CreationQueue}  & 10  & 6\,s/block  & batch \\
4. Upload    & \texttt{UploadQueue}    & 20  & 5$\times$6\,s & concurrent \\
\bottomrule
\end{tabular}
\end{adjustbox}
\end{table}

\noindent
The pipeline uses BullMQ's \texttt{FlowProducer} to create dependent job graphs with \texttt{failParentOnFailure: true}, ensuring that a failure at any stage propagates upward and prevents the parent upload job from executing with incomplete data.

Key implementation details:
\begin{itemize}[nosep]
  \item \textbf{Download stage}: Uses \texttt{yt-dlp} via the \texttt{youtube-dl-exec} wrapper with SOCKS5 proxy routing. A random pre-download sleep of 0--30 seconds introduces behavioral variance. Maximum video size is capped at 15{,}000\,MB (15\,GB) and duration at 10{,}800\,s (3 hours).
  \item \textbf{Metadata stage}: Computes blake3~\cite{blake3} content hashes for integrity verification and extracts resolution, codec, and duration metadata via \texttt{ffprobe}. Timeout is 30 minutes per video.
  \item \textbf{Creation stage}: Batches up to 10 video creation extrinsics into a single Substrate batch call. Uses auto-renewing locks (60\,s duration, 30\,s renewal) to prevent BullMQ stall detection during long batch operations. Application action nonces are managed sequentially within each batch.
  \item \textbf{Upload stage}: Uploads video and thumbnail assets to randomly-selected active Colossus storage nodes with up to 5 retry attempts and 6-second inter-attempt sleep. Sets \texttt{maxRedirects: 0} to prevent HTTP client memory buffering of large payloads.
\end{itemize}

\subsection{Video State Machine}

\begin{figure}[h]
\centering
\begin{tikzpicture}[
  node distance=0.6cm,
  state/.style={draw, rounded corners=2pt, minimum width=1.8cm, minimum height=0.45cm, font=\scriptsize, fill=#1!15, align=center},
  state/.default=blue,
  arr/.style={-{Stealth[length=2mm]}, thick, font=\tiny},
  fail/.style={-{Stealth[length=2mm]}, thick, dashed, red!70!black, font=\tiny},
]
\node[state=gray] (new) at (0,0) {\texttt{New}};
\node[state] (creating) at (0,-0.9) {\texttt{CreatingVideo}};
\node[state] (created) at (0,-1.8) {\texttt{VideoCreated}};
\node[state] (uploading) at (0,-2.7) {\texttt{UploadStarted}};
\node[state=green] (success) at (0,-3.6) {\texttt{UploadSucceeded}};

\node[state=red, minimum width=2.2cm] (cfail) at (3,-0.9) {\texttt{CreationFailed}};
\node[state=red, minimum width=2.2cm] (ufail) at (3,-2.7) {\texttt{UploadFailed}};
\node[state=orange, minimum width=2.6cm] (unavail) at (-3,-1.8) {\texttt{VideoUnavailable::*}};

\draw[arr] (new) -- (creating);
\draw[arr] (creating) -- (created);
\draw[arr] (created) -- (uploading);
\draw[arr] (uploading) -- (success);
\draw[fail] (creating) -- (cfail) node[midway, above, black] {fail};
\draw[fail] (uploading) -- (ufail) node[midway, above, black] {fail};
\draw[arr, green!50!black] (cfail) -- ++(0,-0.6) -| (creating) node[near start, right, black] {retry};
\draw[arr, green!50!black] (ufail) -- ++(0,-0.6) -| (uploading) node[near start, right, black] {retry};
\draw[fail] (new) -- (unavail) node[midway, above, black, sloped] {filter};
\end{tikzpicture}
\caption{Video state machine. Processing flows top-to-bottom through four stages. Failed states retry on the next processing cycle. The \texttt{VideoUnavailable} terminal state has 9 variants: \texttt{Deleted}, \texttt{Private}, \texttt{AgeRestricted}, \texttt{MembersOnly}, \texttt{LiveOffline}, \texttt{DownloadTimedOut}, \texttt{EmptyDownload}, \texttt{PostprocessingError}, \texttt{Skipped}.}
\label{fig:state-machine}
\end{figure}
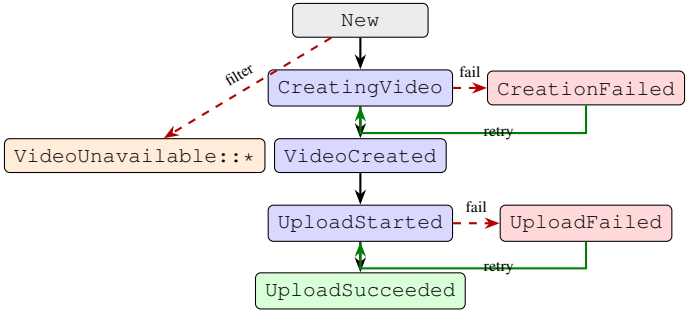

Each video tracked by the system follows a deterministic state machine with 7 processing states and 9 terminal unavailability states (16 total), shown in Figure~\ref{fig:state-machine}.

\subsection{DynamoDB Data Model}

The system uses AWS DynamoDB with 5 tables, all configured with \texttt{PAY\_PER\_REQUEST} billing mode (Section~\ref{sec:dynamodb-incident}). Table~\ref{tab:dynamodb} summarizes the schema design.

\begin{table}[h]
\centering
\caption{DynamoDB Schema Design (5 Tables)}
\label{tab:dynamodb}
\footnotesize
\begin{adjustbox}{max width=\columnwidth}
\begin{tabular}{@{}p{1.8cm}p{1.2cm}p{0.8cm}p{3cm}@{}}
\toprule
\textbf{Table} & \textbf{Hash Key} & \textbf{Range} & \textbf{GSIs} \\
\midrule
\texttt{channels} & \texttt{userId} & \texttt{id} & \texttt{joystreamChannelId}, \texttt{referrerChannelId}, \texttt{phantomKey} \\
\texttt{videos} & \texttt{channelId} & \texttt{id} & \texttt{state-publishedAt} \\
\texttt{users} & \texttt{id} & --- & --- \\
\texttt{stats} & \texttt{partition} & \texttt{date} & --- \\
\texttt{whitelist} & \texttt{handle} & --- & --- \\
\bottomrule
\end{tabular}
\end{adjustbox}
\end{table}

All database operations are serialized through \texttt{AsyncLock} with \texttt{maxPending} set to \texttt{Number.MAX\_SAFE\_IN\-TEGER}, introduced after production race conditions caused data corruption in concurrent write scenarios.

\subsection{Priority Scheduling Algorithm}
\label{sec:priority}

Videos are prioritized for processing using a multi-factor scoring function that balances freshness, channel tier, backlog fairness, and temporal recency. Algorithm~\ref{alg:priority} formalizes the priority computation.

\begin{algorithm}[h]
\caption{Video Priority Computation}
\label{alg:priority}
\footnotesize
\begin{algorithmic}[1]
\Require Video $v$ with \texttt{publishedAt}, \texttt{duration}, channel $c$ with \texttt{tier}, \texttt{backlogPct}
\Ensure BullMQ priority $p \in [0, 2{,}097{,}152]$ (lower = higher priority)
\State $\text{sudo} \gets 10$ \Comment{Default base priority}
\If{$v$ is new \textbf{and} $v.\text{duration} > 300$\,s}
  \State $\text{sudo} \gets \text{sudo} + 20$ \Comment{Fresh content bonus}
\EndIf
\If{$c.\text{tier} \in \{\text{Silver}, \text{Gold}, \text{Diamond}\}$}
  \State $\text{sudo} \gets \text{sudo} + 20$ \Comment{Higher-tier bonus}
\EndIf
\State $\text{recency} \gets v.\text{publishedAt} - 946{,}684{,}800$ \Comment{Seconds since Y2K}
\State $\text{score} \gets c.\text{backlogPct} \times 1000 + \text{sudo} \times 2000 + \text{recency}$
\State $\text{maxScore} \gets 100 \times 2000 + 1000 + 100$
\State $p \gets 2{,}097{,}152 - \lfloor (\text{score} / \text{maxScore}) \times 2{,}097{,}152 \rfloor$
\State \Return $\max(0, p)$
\end{algorithmic}
\end{algorithm}

The constant $2{,}097{,}152$ ($2^{21}$) is BullMQ's maximum priority value. Normalization by \texttt{maxScore} ($= 201{,}100$) ensures the priority maps to the full BullMQ range. Priorities are recalculated periodically, for example when a channel's tier is upgraded.

\subsection{YPP Tier System}

The YouTube Partner Program uses a four-tier structure that governs both creator rewards and system resource allocation. Table~\ref{tab:tiers} shows the production configuration.

\begin{table}[h]
\centering
\caption{YPP Tier Configuration (Production Values)}
\label{tab:tiers}
\begin{adjustbox}{max width=\columnwidth}
\begin{tabular}{@{}lrrcl@{}}
\toprule
\textbf{Tier} & \textbf{Video Cap} & \textbf{Size Cap} & \textbf{Referral} & \textbf{Priority} \\
\midrule
Bronze   & 5       & 1\,GB     & 2    & Default \\
Silver   & 100     & 10\,GB    & 25   & +20 \\
Gold     & 250     & 100\,GB   & 50   & +20 \\
Diamond  & 1{,}000 & 1\,TB     & 100  & +20 \\
\bottomrule
\end{tabular}
\end{adjustbox}
\end{table}

\noindent
Creator onboarding requirements: $\geq$50 subscribers, $\geq$2 videos, channel age $\geq$720 hours (30 days).

\subsection{Authentication and Onboarding Flow}
\label{sec:authentication}

The authentication architecture evolved through two fundamentally different designs, each reflecting the system's changing relationship with YouTube's OAuth infrastructure.

\subsubsection{Phase 1: Google OAuth2 Flow (v1.0--v3.7)}

The original onboarding flow relied on Google's OAuth2 protocol with \texttt{youtube.readonly}, \texttt{userinfo.profile}, and \texttt{userinfo.email} scopes:

\begin{enumerate}[nosep]
  \item The Atlas frontend redirects the creator to Google's OAuth consent screen
  \item Creator authorizes access; Google redirects back with an authorization code
  \item Backend exchanges the code for access and refresh tokens via \texttt{OAuth2Client.getToken()}
  \item Backend extracts the Google user ID (\texttt{sub}) and email from the access token
  \item Channel validation: subscriber count $\geq 50$, video count $\geq 2$, channel age $\geq 720$\,hours
  \item Whitelist exemption check via the \texttt{whitelistChannels} DynamoDB table
  \item User record saved to DynamoDB with tokens, email, and Google ID
  \item Channel registration: validates Joystream channel exists, is not already connected, and referrer is not self-referral
  \item Authorization code replaced with \texttt{randomBytes(10).toString('hex')} to prevent replay attacks
  \item If re-signup after opt-out, the previous tier (\texttt{preOptOutStatus}) is restored
\end{enumerate}

This design stored 7 OAuth-related fields across the \texttt{users} and \texttt{channels} tables (\texttt{accessToken}, \texttt{refreshToken}, \texttt{authorizationCode}, \texttt{email}, \texttt{googleId}, \texttt{userAccessToken}, \texttt{userRefreshToken}), creating a substantial attack surface for the OAuth token lifecycle defense described in Section~\ref{sec:yt-defenses}.

\subsubsection{Phase 2: Video-Based Verification (v4.0+)}

After the OAuth mass-expiration incident (Section~\ref{sec:oauth-incident}), the entire OAuth dependency was eliminated. The replacement protocol requires no Google API interaction:

\begin{enumerate}[nosep]
  \item Creator uploads an unlisted video with prescribed title (``I want to be in YPP'') to their YouTube channel
  \item Creator submits the video URL to the signup endpoint
  \item Backend verifies through the self-hosted Operational API container: video exists, is unlisted, title matches the verification phrase, and channel meets onboarding requirements
  \item On success, a Joystream membership is created and the channel enrollment is recorded with \texttt{youtubeVideoUrl} as the sole verification artifact
\end{enumerate}

This design eliminates \emph{all} Google OAuth dependencies: no API keys, no token exchange, no refresh token lifecycle, no quota consumption. The 7 OAuth-related fields were removed from the schema (Table~\ref{tab:schema-cleanup}), along with corresponding code in the API controllers, repository layer, startup reconciliation, and Pulumi IaC templates.

\section{Deployment Architecture and Infrastructure}
\label{sec:deployment}

\subsection{Containerized Service Topology}

The production deployment consists of four Docker containers orchestrated via \texttt{docker-compose}, connected through a dual-stack network configuration:

\begin{table}[h]
\centering
\caption{Docker Container Topology (Production)}
\label{tab:docker}
\footnotesize
\begin{adjustbox}{max width=\columnwidth}
\begin{tabular}{@{}p{2.4cm}p{1.2cm}p{1.5cm}p{2cm}@{}}
\toprule
\textbf{Container} & \textbf{Image} & \textbf{Ports} & \textbf{Volumes} \\
\midrule
\texttt{httpApi}   & Custom   & API port (host-bound \texttt{127.0.0.1}) & \texttt{logs/}, ADC key \\
\texttt{sync}      & Custom   & None (internal)                          & \texttt{logs/}, \texttt{data/}, ADC key \\
\texttt{redis}     & \texttt{redis:7.2.1} & Redis port (host-bound) & \texttt{redis-data} \\
\texttt{yt-op-api} & OpsAPI   & API port (internal)                       & --- \\
\bottomrule
\end{tabular}
\end{adjustbox}
\end{table}

\noindent
\textbf{Network configuration}: The \texttt{youtube-synch} bridge network uses IPv4 subnet \texttt{172.20.0.0/24} with a dedicated IPv6 subnet (\texttt{2001:db8::/64}) for dual-stack connectivity, enabling proxy routing through both protocol stacks. Port bindings are restricted to \texttt{127.0.0.1} to prevent external access to the API and Redis services.

\textbf{Service decomposition}: The two application containers are built from the same Docker image but started with different \texttt{--service} CLI flags: \texttt{start --service httpApi} for the NestJS REST server (creator onboarding, Swagger docs, BullMQ Board dashboard) and \texttt{start --service sync} for the content processing pipeline. This pattern enables:
\begin{itemize}[nosep]
  \item \textbf{Independent scaling}: API traffic scales with user signups; sync scales with content volume
  \item \textbf{Fault isolation}: API downtime does not halt content processing, and vice versa
  \item \textbf{Different resource profiles}: The sync service is CPU/disk/bandwidth-intensive; the API service is memory/network-bound
\end{itemize}

\subsection{Build and Runtime Stack}

The Docker image is built on \texttt{node:20} and installs several critical system-level dependencies beyond the Node.js application:

\begin{itemize}[nosep]
  \item \texttt{proxychains4}: Transparent SOCKS5 proxy routing for all \texttt{yt-dlp} subprocess invocations
  \item \texttt{awscli}: DynamoDB access via IAM credentials (Application Default Credentials mounted read-only)
  \item Docker-in-Docker capability: Used by Generation~1 proxy (Chisel) for automated EC2 instance restart
  \item \texttt{yt-dlp}: Force-updated at build time via \texttt{youtube-dl-exec} post-install script to ensure the latest YouTube extractor patches
\end{itemize}

The startup script checks the \texttt{TELEMETRY\_ENABLED} environment variable; if set to \texttt{yes}, the application is launched with \texttt{node --require ./opentelemetry/index.js} to enable OpenTelemetry auto-instrumentation before any application code executes.

\subsection{Infrastructure-as-Code}

DynamoDB tables are provisioned via Pulumi IaC templates, which were updated after the throughput incident (Section~\ref{sec:dynamodb-incident}) to default to \texttt{PAY\_PER\_REQUEST} billing. The IaC templates define all 5 tables with their hash/range keys, GSI configurations, and projection types, ensuring reproducible deployment across environments.

Redis is configured with \texttt{maxmemory-policy: noeviction} to prevent BullMQ job data loss under memory pressure---a critical safety property since Redis serves as the ephemeral job queue backing store that is fully reconstructed from DynamoDB on each restart.

\subsection{Scalability Mechanisms}
\label{sec:scalability}

Scaling to 10{,}000+ channels requires careful resource allocation across multiple bottleneck dimensions simultaneously---YouTube's detection sensitivity, blockchain throughput, storage node capacity, and local disk space.

\textbf{Tiered resource allocation}: The YPP tier system (Table~\ref{tab:tiers}) functions as a resource allocator, not merely a reward structure. Bronze channels (5 videos, 1\,GB) consume minimal pipeline capacity, while Diamond channels (1{,}000 videos, 1\,TB) require committed resources. This creates a natural admission control mechanism where most channels (Bronze) occupy bounded space.

\textbf{Polling batch sizing}: The \texttt{YoutubePollingService} processes channels in carefully-tuned batch sizes: 50 channels per batch for stats fetching (matching the YouTube Operational API's \texttt{maxResults} parameter), 100 channels per batch for video ingestion, with a 100\,ms inter-batch sleep to prevent self-inflicted rate limiting against both the Operational API and DynamoDB.

\textbf{Download concurrency tuning}: Download concurrency was reduced from 50 to 2---a 25$\times$ reduction---as a direct anti-detection measure (Section~\ref{sec:anti-detection}). This counter-intuitive sacrifice of throughput for survivability reflects the fundamental tradeoff: a system that downloads quickly but gets blocked within hours processes less content than one that downloads slowly but operates continuously.

\textbf{Upload concurrency}: Upload concurrency remains high at 20, as Colossus storage nodes do not implement the sophisticated bot detection that YouTube does. A random storage node is selected via \texttt{lodash.shuffle()} for each upload, distributing load across the storage provider network.

\textbf{Batch transaction optimization}: Blockchain video creation batches 10 extrinsics per Substrate batch call with auto-renewing BullMQ locks (60\,s duration, 30\,s renewal interval) to prevent stall detection during the multi-block confirmation wait. Application-level action nonces are managed sequentially within each batch to prevent nonce collision.

\textbf{DynamoDB on-demand scaling}: After the throughput incident, all tables use \texttt{PAY\_PER\_REQUEST} billing, providing automatic capacity provisioning that scales with channel enrollment without manual tuning.

\subsection{A Throughput Ceiling from First Principles}
\label{sec:ceiling}

The download stage is the binding constraint (concurrency $C_d=2$). Per-video service time is $T_v = 3\cdot\mathbb{E}[U(0,30\text{s})] + t_f = 45\text{s} + t_f$, where the $45$\,s term is the triple pre-download sleep (known by construction, Section~\ref{sec:anti-detection}) and $t_f$ is the proxied fetch time. Single-instance throughput is
\[
  \lambda = \frac{C_d}{T_v} = \frac{2}{45 + t_f}\ \text{videos/s}.
\]
Even in the physically impossible best case $t_f \to 0$, $\lambda_{\max} = 2/45 \approx 3{,}840$ videos/day; for any realistic $t_f \ge 45$\,s, $\lambda \le 1{,}920$ videos/day.

Steady-state demand at target scale is $\approx 10{,}000 \times \tfrac{1}{7} \approx 1{,}430$ new videos/day, atop an initial backlog of $\approx 10{,}000 \times 20 = 200{,}000$ videos. Thus a single instance operates \emph{at or below} steady-state demand and would require $\ge 52$ days to drain the backlog even in the impossible best case (realistically months). By contrast, on-chain creation ($10$ extrinsics per ${\sim}6$\,s block $\Rightarrow$ tens of thousands/day) is far from binding.

This conclusion holds for the \emph{entire} plausible range of $t_f$ and therefore requires no measurement: detection-driven concurrency limits---not blockchain throughput---are the dominant scaling constraint, making priority triage (Algorithm~\ref{alg:priority}) and horizontal scaling structural necessities rather than optimizations.

\subsection{Fault Tolerance Framework}
\label{sec:fault-tolerance}

The fault tolerance design is anchored in a key invariant: \textbf{DynamoDB is the single source of truth; all other state is ephemeral and reconstructable.} Redis job queues, local downloaded files, and in-memory caches are all derived from DynamoDB state and can be fully rebuilt on restart.

\subsubsection{Startup State Reconciliation}

On every service restart, the system executes a four-phase reconstruction:

\begin{enumerate}[nosep]
  \item \textbf{Redis flush}: All Redis data is purged via \texttt{connection.flushall()}, discarding any stale job state from the previous process lifecycle. This aggressive approach eliminates an entire category of inconsistency bugs at the cost of re-queueing work.

  \item \textbf{Download resolution}: \texttt{ContentDownloadService.start()} scans the \texttt{downloadsDir} and \texttt{THUMBNAILS\_SUBDIR} on disk, re-populating the \texttt{downloadedVideoAssetPaths} map and recalculating \texttt{usedSpace}. Videos already downloaded are not re-downloaded.

  \item \textbf{Creation consistency}: \texttt{ensureContentStateConsistency()} queries DynamoDB for all videos in the \texttt{CreatingVideo} state, then cross-references each against the Joystream blockchain via Orion (GraphQL). If the video exists on-chain, DynamoDB is advanced to \texttt{VideoCreated}; if not, the state is reset to \texttt{New} for re-processing. This reconciles the gap between DynamoDB and blockchain state that occurs when the service crashes between extrinsic submission and state recording.

  \item \textbf{Upload consistency}: \texttt{ensureUploadStateConsistency()} similarly queries all videos in the \texttt{UploadStarted} state. For each, it checks whether both the media asset and thumbnail have been accepted by the storage provider (via GraphQL field \texttt{isAccepted}). Accepted uploads advance to \texttt{UploadSucceeded}; others reset to \texttt{UploadFailed} for retry.
\end{enumerate}

\subsubsection{Write-Ahead Log Pattern}

The pre-commit WAL pattern prevents the most critical failure mode---duplicate on-chain video objects:

\begin{enumerate}[nosep]
  \item Before submitting any blockchain extrinsic, transition video state: \texttt{New} $\rightarrow$ \texttt{CreatingVideo} in DynamoDB
  \item Submit the Substrate batch extrinsic (may require multiple blocks to finalize)
  \item On confirmed success, advance to \texttt{VideoCreated} with the on-chain video ID and asset IDs
  \item On failure or crash: startup reconciliation (step 3 above) resolves the ambiguity
\end{enumerate}

\subsubsection{Concurrency Serialization}

All DynamoDB operations are serialized through \texttt{AsyncLock} instances at the repository level, each with \texttt{maxPending: Number.MAX\_SAFE\_INTEGER}. Five distinct lock domains prevent concurrent write corruption:

\begin{itemize}[nosep]
  \item \textbf{Channel lock} (\texttt{`channel'}): Serializes all channel CRUD including \texttt{save()}, \texttt{query()}, \texttt{batchSave()}, and \texttt{delete()} operations. The lock can be disabled via constructor parameter for non-critical read-only paths.
  \item \textbf{User lock} (\texttt{`user'}): Serializes user record operations, critical during concurrent signup requests.
  \item \textbf{Whitelist lock}: Serializes whitelist exemption checks.
  \item \textbf{Queue locks}: The \texttt{PriorityJobQueue} uses a dual-lock structure---\texttt{RECALCULATE\_PRIORITY\_LOCK\_KEY} per queue (serializes concurrent processing count and priority recalculation) and \texttt{BATCH\_LOCK\_KEY} (prevents concurrent batch execution). Both locks must be acquired for batch processing, preventing races between batch execution and priority recalculation.
  \item \textbf{Proxy lock} (\texttt{`proxy\_bind'}): Serializes proxy assignment with \texttt{maxPending: proxiesNum $\times$ 10}, ensuring concurrent downloads never receive the same proxy endpoint.
\end{itemize}

\subsubsection{Proxy Failover}

When a download encounters YouTube's ``Sign in to confirm you're not a bot'' response, the proxy is reported as faulty via \texttt{reportFaultyProxy(url)}. The \texttt{Socks5ProxyService} maintains a \texttt{NodeCache} with \texttt{stdTTL: 14{,}400} seconds (4 hours), automatically re-enabling the proxy after the exclusion period. If no proxies are available, download jobs spin-wait with a configurable interval rather than failing, preserving the job in the queue for eventual processing.

\section{System Evolution: An Arms Race with YouTube}
\label{sec:evolution}

\begin{figure*}[t]
\centering
\begin{tikzpicture}[
  phase/.style={draw, rounded corners=3pt, minimum width=2.2cm, minimum height=1.4cm, font=\scriptsize, align=center, fill=#1!12, text width=2.5cm},
  arr/.style={-{Stealth[length=3mm]}, ultra thick, gray!60},
]

\node[phase=phaseA] (p0) at (0,0) {\textbf{Phase 0}\\Exploration\\11 PRs\\Jun'21--Aug'22};
\node[phase=phaseB] (p1) at (3.2,0) {\textbf{Phase 1}\\Production Proto\\$\sim$30 PRs\\Sep--Dec'22};
\node[phase=phaseC] (p2) at (6.4,0) {\textbf{Phase 2}\\Hardening/QA\\$\sim$25 PRs\\Jan--May'23};
\node[phase=phaseD] (p3) at (9.6,0) {\textbf{Phase 3}\\BullMQ/YPP\,2.0\\$\sim$20 PRs\\Jun--Oct'23};
\node[phase=phaseE] (p4) at (12.8,0) {\textbf{Phase 4}\\Proxy/Split\\$\sim$15 PRs\\Nov'23--Feb'24};
\node[phase=phaseF] (p5) at (16,0) {\textbf{Phase 5}\\API-Free\\$\sim$29 PRs\\Mar'24--Present};

\draw[arr] (p0) -- (p1);
\draw[arr] (p1) -- (p2);
\draw[arr] (p2) -- (p3);
\draw[arr] (p3) -- (p4);
\draw[arr] (p4) -- (p5);

\node[font=\scriptsize, red!70!black, below=0.05cm of p2, text width=2.5cm, align=center] {28 duplicate videos\\DynamoDB incident};
\node[font=\scriptsize, red!70!black, below=0.05cm of p4, text width=2.5cm, align=center] {YouTube IP\\blocking begins};
\node[font=\scriptsize, red!70!black, below=0.05cm of p5, text width=2.5cm, align=center] {10K+ channels\\mass opt-out};
\end{tikzpicture}
\caption{Development phases with pull request counts and critical incidents. Each phase transition was driven by production requirements or platform policy changes.}
\label{fig:timeline}
\end{figure*}
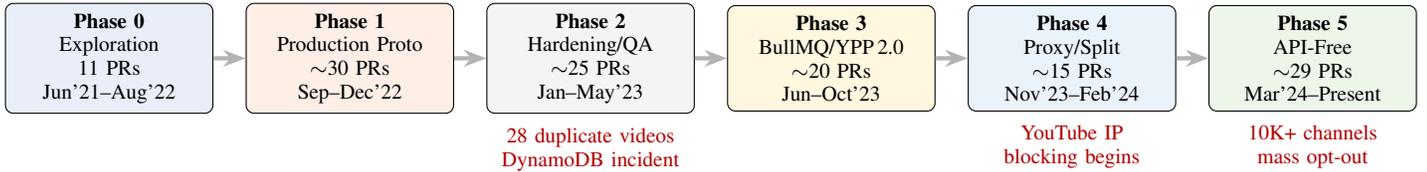

YouTube-Synch's evolution is best understood not as a conventional software maturation story, but as a sustained arms race with a platform backed by billions of dollars of infrastructure investment. Each phase represents either (a) a proactive attempt to circumvent a known YouTube defense layer, or (b) a reactive architectural overhaul forced by an unexpected interaction between YouTube's defenses and the system's current design. Figure~\ref{fig:timeline} illustrates the phase transitions with their forcing functions.

\subsection{Phase 0: Exploration---Learning YouTube's Constraints (Jun 2021 -- Aug 2022)}

The project originated from a requirements specification targeting web-based self-service signup with horizontal scaling to 10 million channels averaging 20 videos each plus 1 new video per week. LBRY's \texttt{ytsync} tool was referenced as prior art, but notably LBRY operated \emph{without} creator authorization---a design decision that would have exposed YouTube-Synch to both legal risk and the full force of YouTube's anti-scraping defenses.

During this period, 11 foundational pull requests explored multiple architectural patterns, each encountering a different YouTube or infrastructure constraint:

\textbf{Discarded architectures:}
\begin{itemize}[nosep]
  \item \textbf{AWS Lambda}: Abandoned due to the 15-minute execution timeout and 10\,GB memory limit---insufficient for downloading long-form video content.
  \item \textbf{Kubernetes orchestration}: Proposed for download worker scaling but simplified to a monolithic service, reducing operational complexity.
  \item \textbf{YouTube Push Notifications} via PubSubHubbub: Implemented but abandoned in favor of polling after encountering undocumented scaling limits. Polling was identified as the more reliable primitive for channel monitoring at scale.
  \item \textbf{PostgreSQL/RDS}: Initially proposed for the database layer but replaced with DynamoDB for its AWS-native auto-scaling and pay-per-request billing model.
  \item \textbf{S3 intermediate storage}: Considered for download staging between pipeline stages but removed in favor of local disk to reduce latency and cost.
\end{itemize}

The phase concluded with an NestJS~\cite{nestjs} migration from ad-hoc Express endpoints to a structured service/controller/module architecture, and establishment of the repository pattern for DynamoDB access.

\subsection{Phase 1: First Contact with YouTube's Quota Wall (Sep -- Dec 2022)}

Phase 1 represents the system's first encounter with YouTube's API quota as a hard operational constraint. The daily budget of 10{,}000 units was immediately revealed to be insufficient for the intended scale, forcing a 5\%/95\% split between signup (500 units) and sync (9{,}500 units) operations. A pivotal 136\,KB pull request replaced the exploratory codebase with a production-grade architecture. Approximately 30 pull requests in this phase established:

\begin{itemize}[nosep]
  \item Channel opt-out, suspension states, and duplicate verification handling
  \item Video upload logic, thumbnail syncing, video categories, and duration metadata
  \item YouTube API quota rationing: 5\% for signup, 95\% for sync operations, monitored via Google Cloud Monitoring API. Even with this allocation, the system could only poll ${\sim}$2{,}500 channels per day---a ceiling that would force the migration away from the API entirely in Phase~4.
  \item The multi-factor priority scheduling algorithm (Section~\ref{sec:priority}), critical for maximizing the value extracted from limited processing capacity
  \item OAuth redirect URL handling for Atlas frontend integration---the beginning of an OAuth dependency that would later interact catastrophically with the API migration
  \item Content filtering: skip private, unlisted, and age-restricted videos
\end{itemize}

\subsection{Phase 2: Infrastructure Hardening---Discovering Internal Fault Lines (Jan -- May 2023)}

While Phases 0--1 focused on YouTube's external defenses, Phase 2 revealed that the system's own infrastructure contained critical fault lines. A structured fault tolerance testing campaign was conducted against all external dependency failures, covering four failure categories: RPC Node, Query Node, Storage Node, and Google API. The campaign's methodology---mocking each dependency as unresponsive, validating error handling, and documenting recovery behavior---uncovered 6 bugs including missing attribution fields that would have broken content provenance tracking. Table~\ref{tab:fault-tolerance} summarizes the results.

\begin{table}[h]
\centering
\caption{Fault Tolerance QA Results}
\label{tab:fault-tolerance}
\begin{adjustbox}{max width=\columnwidth}
\begin{tabular}{@{}lcc@{}}
\toprule
\textbf{Failed Dependency} & \textbf{Bugs Found} & \textbf{Auto-Recovery} \\
\midrule
Joystream RPC Node    & 1 & Yes (retry) \\
Query Node / Orion    & 2 & Yes (retry) \\
Colossus Storage Node & 1 & Yes (retry) \\
Google YouTube API    & 2 & Partial \\
\midrule
\textbf{Total}        & \textbf{6} & --- \\
\bottomrule
\end{tabular}
\end{adjustbox}
\end{table}

Specific bugs discovered:
\begin{enumerate}[nosep]
  \item Missing \texttt{ytVideoId} and \texttt{entryApp} attribution fields on synced videos
  \item Deleted YouTube channels remaining as ``Unverified'' instead of transitioning to ``OptedOut''
  \item Error codes not included in API error responses
  \item Channel opt-out flow failures for revoked Google permissions
\end{enumerate}

This phase also saw the critical DynamoDB incident (Section~\ref{sec:dynamodb-incident}).

\subsection{Phase 3: BullMQ Migration---Scaling the Pipeline to Match YouTube's Volume (Jun -- Oct 2023)}
\label{sec:bullmq}

The original \texttt{better-queue} library became a scaling bottleneck as channel enrollment grew. Without persistent job queues, priority scheduling, or DAG-based flow control, the system could not efficiently process the volume of videos being ingested from 10{,}000+ channels. Version 3.0.0 introduced BullMQ with a 231\,KB pull request---the largest feature merge in the project's history. Table~\ref{tab:queue-migration} compares the two queue systems.

\begin{table}[h]
\centering
\caption{Queue System Migration: better-queue $\rightarrow$ BullMQ}
\label{tab:queue-migration}
\begin{adjustbox}{max width=\columnwidth}
\begin{tabular}{@{}lcc@{}}
\toprule
\textbf{Feature} & \textbf{better-queue} & \textbf{BullMQ 4.11} \\
\midrule
Persistence          & In-memory     & Redis-backed \\
Flow/DAG jobs        & No            & Yes \\
Priority scheduling  & No            & Yes (0--$2^{21}$) \\
Dashboard            & No            & Bull Board \\
Batch processing     & No            & Custom impl. \\
Stall detection      & No            & Yes \\
Auto-renewing locks  & No            & Yes (60\,s) \\
\bottomrule
\end{tabular}
\end{adjustbox}
\end{table}

The migration required adding Redis 7.2.1 as a new infrastructure component with \texttt{maxmemory-policy: noeviction} to prevent BullMQ data loss under memory pressure.

YPP 2.0 introduced the four-tier system (Bronze/Silver/Gold/Diamond) replacing the original subscriber-count-based tiers, along with per-channel sync limits, batch video creation transactions (\texttt{createVideoTxBatchSize}: 10), and enhanced onboarding requirements.

\subsection{Phase 4: Breaching the IP Wall---Proxy Infrastructure and Service Decomposition (Nov 2023 -- Feb 2024)}

This phase marks the inflection point where the system's relationship with YouTube transitioned from operating within YouTube's rules (API quota) to operating \emph{around} them (direct scraping with proxy evasion).

After Phase 3's migration from YouTube API to \texttt{yt-dlp} for all data fetching, the system's access pattern changed fundamentally: instead of authenticated API calls distributed across Google's CDN, it now made unauthenticated HTTP requests that were concentrated on YouTube's video serving infrastructure. YouTube's behavioral analysis detected this high-volume scraping pattern and blocked the service's IP address within days. This forced the introduction of proxy infrastructure (Section~\ref{sec:proxy-evolution}), beginning the multi-generational anti-detection arms race.

Simultaneously, version 3.4.0 decomposed the monolith into two independently-deployable Docker services---\texttt{httpApi} and \texttt{sync}---connected via a \texttt{--service} CLI flag. This enabled:
\begin{itemize}[nosep]
  \item Independent scaling (API proportional to user/signup traffic; sync proportional to content volume)
  \item Fault isolation (API downtime does not halt content processing)
  \item Different resource profiles (sync is CPU/disk/bandwidth-intensive; API is memory/network-intensive)
\end{itemize}

OpenTelemetry~\cite{opentelemetry} tracing was integrated in v3.4.0 with batch span processing and OTLP export, providing distributed tracing across the pipeline stages. Version 3.6.0 migrated from the deprecated Query Node to Orion for all GraphQL queries, including the migration from \texttt{WebSocketLink} to \texttt{GraphQLWsLink} for GraphQL subscriptions.

\subsection{Phase 5: Complete Decoupling---Eliminating YouTube's Last Leverage Point (Mar 2024 -- Present)}

The mass opt-out incident (Section~\ref{sec:oauth-incident}) demonstrated the cascading nature of YouTube's defense layers: the migration away from the API (Phase 4) had rendered OAuth tokens unused, activating Google's 6-month inactive token expiration policy---a defense layer that was invisible until triggered. Version 4.0.0 responded by eliminating Google OAuth entirely, removing YouTube's last remaining leverage point over the system.

\begin{table}[h]
\centering
\caption{Schema Fields Removed in v4.0 API-Free Migration}
\label{tab:schema-cleanup}
\begin{adjustbox}{max width=\columnwidth}
\begin{tabular}{@{}ll@{}}
\toprule
\textbf{Table} & \textbf{Fields Removed} \\
\midrule
\texttt{channels} & \texttt{userAccessToken}, \texttt{userRefreshToken}, \\
                   & \texttt{email}, \texttt{uploadsPlaylistId} \\
\texttt{users}    & \texttt{googleId}, \texttt{authorizationCode}, \\
                   & \texttt{accessToken}, \texttt{refreshToken}, \texttt{email} \\
\bottomrule
\end{tabular}
\end{adjustbox}
\end{table}

A maintainership transition occurred, with new contributors adding 14 pull requests focused on operational improvements: \texttt{proxychains4} proxy migration, a queue cleanup CLI command, pre-download format/size verification, and emergency operational controls (\texttt{disableNewSignUps} configuration).

\textbf{Defense layer decoupling summary}: Table~\ref{tab:defense-decoupling} summarizes the cumulative effect of each phase on YouTube's defense layers, showing the progressive elimination of YouTube's control surfaces.

\begin{table}[h]
\centering
\caption{YouTube Defense Layer Status by Phase}
\label{tab:defense-decoupling}
\footnotesize
\begin{adjustbox}{max width=\columnwidth}
\begin{tabular}{@{}lcccc@{}}
\toprule
\textbf{Defense Layer} & \textbf{P0--P1} & \textbf{P2--P3} & \textbf{P4} & \textbf{P5} \\
\midrule
API Quota        & Active & Active   & Bypassed   & Eliminated \\
OAuth Tokens     & Active & Active   & Active     & Eliminated \\
IP Blocking      & N/A    & N/A      & \textbf{Active} & Mitigated \\
Bot Detection    & N/A    & N/A      & \textbf{Active} & Mitigated \\
\bottomrule
\end{tabular}
\end{adjustbox}
\end{table}

\noindent
By Phase~5, the only remaining active YouTube defense layers are IP blocking and bot detection---and these are addressed through the proxy and behavioral variance infrastructure rather than eliminated. This reflects a fundamental asymmetry: authentication-based defenses can be fully decoupled from, but access-pattern-based defenses require ongoing operational countermeasures.

\section{Production Challenges: When YouTube's Defenses Interact with System Architecture}
\label{sec:challenges}

\subsection{YouTube API Quota Evolution: From Rationing to Abandonment}
\label{sec:quota-evolution}

\begin{figure}[h]
\centering
\begin{tikzpicture}[
  node distance=0.35cm,
  stage/.style={draw, rounded corners=3pt, minimum width=6.5cm, minimum height=0.6cm, font=\scriptsize, fill=#1!12, align=left},
  arr/.style={-{Stealth[length=2mm]}, thick},
]
\node[stage=red] (s1) at (0,0) {v1.0--v1.4: Full API (10K units/day budget)};
\node[stage=orange, below=of s1] (s2) {v3.2: yt-dlp for video metadata ($\downarrow$50\% API)};
\node[stage=yellow, below=of s2] (s3) {v3.3: yt-dlp for channel details ($\downarrow$90\% API)};
\node[stage=green, below=of s3] (s4) {v4.0: Operational API (0\% YouTube API)};
\draw[arr] (s1) -- (s2) node[midway, right, font=\tiny] {Quota exhaustion};
\draw[arr] (s2) -- (s3) node[midway, right, font=\tiny] {Continued pressure};
\draw[arr] (s3) -- (s4) node[midway, right, font=\tiny] {OAuth mass-expiry};
\end{tikzpicture}
\caption{YouTube API dependency reduction timeline. Each transition was forced by an operational constraint---quota exhaustion, continued budget pressure, or the OAuth mass-expiration incident.}
\label{fig:quota-evolution}
\end{figure}
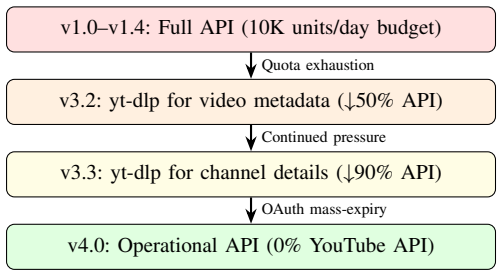

The system's relationship with YouTube's API is a story of progressive escape from a resource ceiling that could never be raised. Each optimization extracted more value from the fixed 10{,}000 unit/day budget---until the fundamental insufficiency of the quota forced a complete architectural departure. Figure~\ref{fig:quota-evolution} traces this progression.

\textbf{v1.0--v1.4 quota rationing}: The daily budget of 10{,}000 units was split between signup (500 units, 5\%) and sync (9{,}500 units, 95\%) services. Even with this allocation, the system could only poll ${\sim}$2{,}500 channels per day at 1 API call per channel per day---insufficient for 10{,}000+ enrolled channels.

\textbf{v3.2 breakthrough}: Migrating video metadata fetching from YouTube API to \texttt{yt-dlp} reduced API dependency by approximately 50\%, as video detail retrieval was the largest quota consumer.

\textbf{v4.0 complete decoupling}: The self-hosted YouTube Operational API container eliminated all YouTube API quota consumption, making the system's scalability independent of Google's quota policies.

\subsection{DynamoDB Throughput Incident}
\label{sec:dynamodb-incident}

\textbf{Timeline}: A channel with 48 videos signed up for YPP. After successful on-chain video object creation, the subsequent DynamoDB state update writes exceeded the provisioned capacity of 1 read unit / 1 write unit.

\textbf{Impact}: \texttt{Provisioned\-Throughput\-Exceeded\-Exception} caused 28 videos to be silently skipped in the state update. On the next processing cycle, the system did not find these videos in a ``created'' state and re-created them, producing \textbf{28 duplicate video objects on the Joystream blockchain}.

\textbf{Root cause}: Production DynamoDB tables were provisioned with minimum capacity (1 RCU / 1 WCU). Community load testing was performed against local DynamoDB (which does not enforce capacity limits), creating a gap in test coverage.

\textbf{Resolution} (3 changes):
\begin{enumerate}[nosep]
  \item Switched all tables to \texttt{PAY\_PER\_REQUEST} billing (on-demand auto-scaling)
  \item Updated Pulumi IaC template defaults to prevent recurrence
  \item Introduced a \textbf{pre-commit WAL pattern}: video state transitions from \texttt{New} $\rightarrow$ \texttt{CreatingVideo} \emph{before} submitting the on-chain extrinsic; on startup, \texttt{ensureStateConsistency()} cross-references \texttt{CreatingVideo} entries against on-chain state to reconcile
\end{enumerate}

\subsection{OAuth Token Mass-Expiration: Cascading Defense Layer Interaction}
\label{sec:oauth-incident}

This incident is the clearest illustration of how YouTube's defense layers interact nonlinearly---a change made to circumvent one defense (API quota) activated a different defense (token lifecycle) months later.

\textbf{Context}: After migrating from YouTube API to \texttt{yt-dlp} for channel data fetching (${\sim}$v3.3), OAuth refresh tokens stored in DynamoDB were no longer being used for any API calls. Google's OAuth2 policy specifies that refresh tokens expire after 6 months of inactivity---a policy designed to limit the window of credential compromise, but which here functioned as an invisible tripwire linked to the API migration.

\textbf{Trigger}: A later code change re-introduced a YouTube API call path that required valid OAuth tokens. Over 10{,}000 channels had tokens that had silently expired.

\textbf{Impact}: The system automatically opted out all channels with invalid tokens, causing \textbf{10{,}000+ channels to lose their YPP enrollment} in a single polling cycle.

\textbf{Emergency response}:
\begin{enumerate}[nosep]
  \item Production hotfix: disabled the automatic opt-out logic for Google permissions errors
  \item Polling interval reduced from 1{,}440 minutes (24 hours) to 2 minutes for faster recovery validation
  \item Proposed channel status reversion using HubSpot CRM versioned field history
\end{enumerate}

\textbf{Architectural consequence}: This incident directly accelerated the v4.0 API-free signup design, completely eliminating the Google OAuth dependency.

\subsection{Proxy Infrastructure: Three Generations of IP Evasion}
\label{sec:proxy-evolution}

YouTube's IP-based blocking forced three generations of proxy architecture, each more sophisticated than the last. The progression reflects YouTube's own adaptation---as the system deployed more capable proxy infrastructure, YouTube's detection became more sensitive, necessitating further countermeasures. Table~\ref{tab:proxy-gen} compares each generation.

\begin{table}[h]
\centering
\caption{Proxy Infrastructure Evolution Across Three Generations}
\label{tab:proxy-gen}
\begin{adjustbox}{max width=\columnwidth}
\begin{tabular}{@{}p{1.5cm}p{1.4cm}p{1.4cm}p{2.2cm}@{}}
\toprule
\textbf{Property} & \textbf{Gen 0} & \textbf{Gen 1} & \textbf{Gen 2} \\
& \textbf{(Direct)} & \textbf{(Chisel)} & \textbf{(proxychains4)} \\
\midrule
Version    & v1.0--3.2 & v3.3--3.7 & v3.8+ \\
Proxies    & 0         & 1 (EC2)   & $N$ (pool) \\
IP rotation & N/A      & EC2 restart & Pool rotation \\
Docker sock & No       & Yes       & No \\
Pre-dl sleep & No      & No        & 0--30\,s \\
Size check  & No       & No        & Yes (15\,GB) \\
\bottomrule
\end{tabular}
\end{adjustbox}
\end{table}

\textbf{Generation 1 --- Chisel (v3.3)}: A Chisel~\cite{chisel} TCP-over-HTTP tunnel connected the YouTube-Synch container to an EC2 proxy instance. IP rotation was achieved by restarting the EC2 instance (which assigns a new public IP). This required mounting the Docker socket into the container for automated restart capability.

\begin{figure}[h]
\centering
\begin{adjustbox}{max width=\columnwidth}
\begin{tikzpicture}[
  node distance=0.4cm,
  box/.style={draw, rounded corners=2pt, minimum height=0.5cm, font=\scriptsize, fill=#1!15, align=center},
  box/.default=blue,
  arr/.style={-{Stealth[length=2mm]}, thick},
]
\node[font=\footnotesize\bfseries] (g1label) at (0,1.5) {Generation 1 (Chisel)};
\node[box] (ytdlp1) at (0,0.7) {yt-dlp};
\node[box=orange] (chisel1) at (2.3,0.7) {Chisel Client};
\node[box=orange] (ec2) at (4.6,0.7) {EC2 / Chisel Srv};
\node[box=red] (youtube1) at (6.9,0.7) {YouTube};
\draw[arr] (ytdlp1) -- (chisel1);
\draw[arr] (chisel1) -- (ec2);
\draw[arr] (ec2) -- (youtube1);

\node[font=\footnotesize\bfseries] (g2label) at (0,-0.3) {Generation 2 (proxychains4)};
\node[box] (ytdlp2) at (0,-1.1) {yt-dlp};
\node[box=green!60!black] (pc4) at (2.3,-1.1) {proxychains4};
\node[box=green!60!black] (pool) at (4.6,-1.1) {SOCKS5 Pool ($N$)};
\node[box=red] (youtube2) at (6.9,-1.1) {YouTube};
\draw[arr] (ytdlp2) -- (pc4);
\draw[arr] (pc4) -- (pool);
\draw[arr] (pool) -- (youtube2);

\end{tikzpicture}
\end{adjustbox}
\caption{Proxy architecture comparison. Generation 1 used a single Chisel tunnel through an EC2 instance. Generation 2 uses proxychains4 with a configurable pool of $N$ SOCKS5 proxy endpoints.}
\label{fig:proxy-arch}
\end{figure}
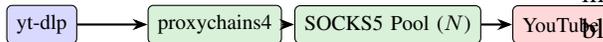

\textbf{Generation 2 --- proxychains4 (v3.8+)}: Replaced Chisel with \texttt{proxychains4}, supporting multiple SOCKS5 proxy URLs configured as an array. The \texttt{Socks5ProxyService} manages proxy assignment using \texttt{AsyncLock} for thread safety, with faulty proxy exclusion for a configurable duration (default: 14{,}400\,s = 4 hours). The \texttt{proxychains4.conf} file is mounted as a Docker volume, enabling proxy reconfiguration without image rebuilds.

\subsection{Queue Pollution and Cleanup}
\label{sec:queue-cleanup}

A polling logic regression allowed invalid videos to enter the processing queue---videos that were missing from YouTube, private, livestreams, age-restricted, or exceeding duration limits. These accumulated as persistently-failing jobs.

\textbf{Impact}: 719 daily size-cap errors before the fix.

\textbf{Solution}: A dedicated \texttt{queueCleanup} CLI command was developed to:
\begin{enumerate}[nosep]
  \item Fetch YouTube metadata for all queued videos (batched in groups of 50)
  \item Categorize by failure reason (Table~\ref{tab:cleanup-categories})
  \item Create JSON backups before deletion
  \item Remove invalid entries from DynamoDB in batches
\end{enumerate}

\begin{table}[h]
\centering
\caption{Queue Cleanup Video Categories}
\label{tab:cleanup-categories}
\begin{adjustbox}{max width=\columnwidth}
\begin{tabular}{@{}ll@{}}
\toprule
\textbf{Category} & \textbf{Action} \\
\midrule
Missing from YouTube   & Delete \\
Private / Members-only & Mark \texttt{VideoUnavailable} \\
Age-restricted         & Mark \texttt{VideoUnavailable} \\
Livestream (live/ended)& Mark \texttt{VideoUnavailable} \\
Exceeds max duration   & Mark \texttt{VideoUnavailable} \\
Exceeds max size       & Mark \texttt{VideoUnavailable} \\
Region-restricted      & Mark \texttt{VideoUnavailable} \\
Missing metadata       & Requeue for retry \\
\bottomrule
\end{tabular}
\end{adjustbox}
\end{table}

Pre-download format and size verification was added to prevent future accumulation, reducing the daily error count from 719 to near-zero.

\subsection{Anti-Detection: Behavioral Variance and Traffic Shaping}
\label{sec:anti-detection}

The system's anti-detection strategy operates on the principle that \emph{survivability trumps throughput}---a system that downloads slowly but operates continuously extracts more content than one that downloads quickly but gets blocked within hours. This section details the layered countermeasures that collectively create a traffic pattern indistinguishable from a small number of human users.

\textbf{1. Triple pre-download sleep}: Each download invokes \texttt{preDownloadSleep()} \emph{three times}---before video format verification (\texttt{checkVideo()}), before thumbnail download (\texttt{downloadThumbnail()}), and before video download (\texttt{downloadVideo()}). Each invocation generates a uniformly random sleep between 0 and 30 seconds (configured as \texttt{preDownloadSleep: \{min: 0, max: 30000\}}), injecting 0--90 seconds of total behavioral variance per video.

\textbf{2. Concurrency reduction}: Download concurrency was reduced 25$\times$ (from 50 to 2) to minimize the system's concurrent connection footprint. Both the download queue and metadata queue share this concurrency limit, meaning at most 4 simultaneous connections to YouTube exist at any time.

\textbf{3. yt-dlp invocation hardening}: Each \texttt{yt-dlp} invocation uses:
\begin{itemize}[nosep]
  \item \texttt{forceIpv4: true} to prevent IPv6 leaks that bypass SOCKS5 proxying
  \item \texttt{retries: 0} to prevent \texttt{yt-dlp}'s internal retry logic from amplifying blocked requests
  \item \texttt{bufferSize: 64K} to control download chunk sizes
  \item Format priority chain: MP4+M4A $\rightarrow$ WebM $\rightarrow$ best, capped at 1080p and 15\,GB
  \item SOCKS5 proxy URL passed per-invocation to \texttt{yt-dlp}'s \texttt{--proxy} flag
\end{itemize}

\textbf{4. Proxy-level bot detection response}: When YouTube responds with ``Sign in to confirm you're not a bot,'' the system:
\begin{enumerate}[nosep]
  \item Logs the blocked proxy via \texttt{reportFaultyProxy()}
  \item Adds the proxy URL to a \texttt{NodeCache} with a 4-hour TTL (\texttt{stdTTL: 14400})
  \item Subsequent \texttt{bindProxy()} calls exclude faulty proxies via set difference
  \item After TTL expiry, the proxy is automatically re-enabled
\end{enumerate}

\textbf{5. proxychains4 configuration}: The SOCKS5 proxy configuration generated at runtime uses \texttt{strict\_chain} mode (sequential proxy chain), \texttt{quiet\_mode} (suppress logging), and conservative timeouts (\texttt{tcp\_read\_time\_out 15000}, \texttt{tcp\_connect\_time\_out 8000}). The configuration file is mounted as a Docker volume, enabling proxy rotation without container rebuilds.

\textbf{6. Pre-download validation}: Before allocating download resources, \texttt{checkVideo()} verifies the video's availability, size, and duration through a lightweight metadata fetch. Videos exceeding 15\,GB or 3 hours are rejected before consuming proxy bandwidth, reducing unnecessary exposure to YouTube's detection systems.

\subsection{State Consistency and the WAL Pattern}

The pre-commit Write-Ahead Log pattern, described in detail in Section~\ref{sec:fault-tolerance}, prevents the most critical failure mode unique to cross-platform replication: blockchain objects are immutable once created, so a service crash between extrinsic submission and state recording produces irrecoverable duplicate on-chain objects. The pattern's effectiveness was validated by the DynamoDB incident (Section~\ref{sec:dynamodb-incident}), which occurred \emph{before} the WAL was introduced and produced 28 such duplicates. After the WAL's introduction, zero duplicate objects have been created despite numerous service restarts and infrastructure failures.

Redis queue state is \textbf{flushed entirely on startup}, with all jobs reconstructed from DynamoDB state---making DynamoDB the single source of truth and Redis purely ephemeral. This aggressive approach eliminates an entire category of stale-queue bugs at the cost of re-analyzing and re-queueing in-progress work on each restart.

\section{Trust-Minimized Design}
\label{sec:trust}

\subsection{Video-Based Ownership Verification}

Version 4.0 replaced Google OAuth with a trust-minimized ownership verification protocol. The creator proves channel ownership by uploading a specific unlisted video, which the system verifies without any Google API interaction:

\begin{enumerate}[nosep]
  \item Creator uploads an unlisted video with prescribed title (e.g., ``I want to be in YPP'') to their YouTube channel
  \item Creator submits the video URL via the signup endpoint
  \item YouTube-Synch verifies via the self-hosted Operational API:
  \begin{itemize}[nosep]
    \item Video exists and is unlisted
    \item Title matches the expected verification phrase
    \item Channel meets onboarding requirements ($\geq$50 subscribers, $\geq$2 videos, age $\geq$720 hours)
  \end{itemize}
  \item On success, a Joystream membership is created and associated with the channel
\end{enumerate}

This design eliminates all Google OAuth dependencies: no API keys, no token exchange, no refresh token lifecycle management, and no quota consumption.

\subsection{Towards Verifiable Attribution via zkSNARKs}

A forward-looking research direction addresses \textbf{non-interactively verifiable attribution}---proving that a person who uploaded content to Joystream actually owns the corresponding YouTube channel, without trusting the YPP operator.

\textbf{Threat model}: A malicious gateway operator could scrape YouTube content without publisher authorization and claim DAO rewards. Manual council verification does not scale to 10{,}000+ channels.

\textbf{Proposed approach}: zkSNARK-based TLS oracle proofs, where the operator produces a zero-knowledge proof that ``YouTube server $X$ provided an access token for channel $Y$'' with the access token as a private witness. This builds on:
\begin{itemize}[nosep]
  \item DECO~\cite{deco}: Decentralized oracle for TLS sessions
  \item TLS-N~\cite{tlsn}: Non-repudiation evidence for TLS connections
\end{itemize}

This remains an open research direction without implementation.

\section{Observability and Operations}
\label{sec:observability}

\subsection{Logging Architecture}

The system uses Winston~\cite{winston} with three transport layers:

\begin{itemize}[nosep]
  \item \textbf{Console}: Verbose level, real-time operational monitoring
  \item \textbf{File}: Debug level, daily rotation, 50\,MB maximum per file, 30 files (30 days) retention
  \item \textbf{Elasticsearch}: ECS-format structured logging, supporting Kibana dashboards and automated alerting
\end{itemize}

A custom \texttt{pauseFormat} middleware suppresses repeated log messages using blake2 hashing, reducing log volume for recurring transient errors (e.g., storage node timeouts). HTTP middleware captures all request/response pairs, including response bodies for failed requests.

\subsection{Distributed Tracing}

OpenTelemetry auto-instrumentation (v3.4+) provides end-to-end tracing across pipeline stages. The OTLP batch span processor exports traces for visualization. DNS instrumentation is excluded to avoid compatibility issues. Tracing is activated per-container via the \texttt{TELEMETRY\_ENABLED=yes} environment variable.

\subsection{CRM Integration}

The \texttt{ypp-operations} package synchronizes channel data with HubSpot CRM via a daily cron job (11:00 AM UTC). Over 25 CRM-related issues were addressed during development, covering: duplicate contact prevention, missing field propagation, suspended channel status sync, automated payout tracking, and batch payment transaction generation. The package uses LokiJS for ephemeral local caching and PM2 for production process management.

\section{Operational Experience and Analysis}
\label{sec:evaluation}

\subsection{Development Metrics}

Table~\ref{tab:dev-metrics} summarizes the quantitative development history.

\begin{table}[h]
\centering
\caption{Development Metrics Summary}
\label{tab:dev-metrics}
\begin{adjustbox}{max width=\columnwidth}
\begin{tabular}{@{}lr@{}}
\toprule
\textbf{Metric} & \textbf{Value} \\
\midrule
Total merged pull requests      & 144 \\
Open pull requests               & 2 \\
Closed unmerged pull requests    & 5 \\
Total issues (open + closed)     & 203 (34 + 169) \\
Release versions                 & 15 (v1.0.0 -- v4.0.0) \\
Active development period        & 3.5 years \\
Production dependencies          & 63 npm packages \\
Dev dependencies                 & 19 npm packages \\
DynamoDB tables                  & 5 \\
BullMQ queues                    & 4 \\
Docker containers (production)   & 4 \\
Video states (total)             & 16 (7 processing + 9 terminal) \\
Channel statuses                 & 10 \\
\bottomrule
\end{tabular}
\end{adjustbox}
\end{table}

\subsection{Contributor Analysis}

Table~\ref{tab:contributors} shows the distribution of merged pull requests across contributors, revealing the typical open-source pattern of one dominant contributor with specialized support.

\begin{table}[h]
\centering
\caption{Contributor Pull Request Distribution}
\label{tab:contributors}
\begin{adjustbox}{max width=\columnwidth}
\begin{tabular}{@{}llr@{}}
\toprule
\textbf{Contributor} & \textbf{Primary Role} & \textbf{Merged PRs} \\
\midrule
Lead developer    & Architecture, pipeline, all versions  & $\sim$80 \\
Initial developer & Foundation, NestJS, DB layer          & 11 \\
Late maintainer   & Proxy, queue cleanup, operations      & 14 \\
Infra engineer    & Production hotfixes, polling tuning   & 3 \\
Frontend dev      & Induction requirements, referrals     & 1 \\
\midrule
\textbf{Total}    &                                       & \textbf{144} \\
\bottomrule
\end{tabular}
\end{adjustbox}
\end{table}

\subsection{Incident Impact Quantification}

Table~\ref{tab:incidents} quantifies the three major production incidents and their resolution effectiveness.

\begin{table}[h]
\centering
\caption{Production Incidents: Impact and Resolution}
\label{tab:incidents}
\begin{adjustbox}{max width=\columnwidth}
\begin{tabular}{@{}p{2.5cm}rrl@{}}
\toprule
\textbf{Incident} & \textbf{Impact} & \textbf{Fix} & \textbf{Prevention} \\
\midrule
DynamoDB throughput exceeded & 28 duplicate videos & v1.2 & PAY\_PER\_REQUEST + WAL \\
OAuth mass-expiration & 10{,}000+ channels opted out & Hotfix & v4.0 API-free signup \\
Queue pollution & 719 errors/day & v3.8+ & Pre-download checks \\
\bottomrule
\end{tabular}
\end{adjustbox}
\end{table}

\subsection{Architecture Complexity Evolution}

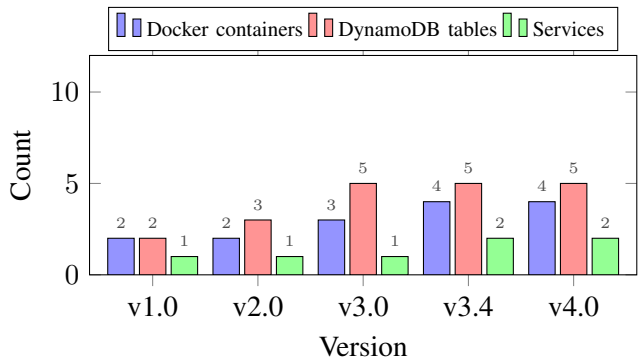
\begin{figure}[h]
\centering
\begin{tikzpicture}
\begin{axis}[
  ybar,
  bar width=10pt,
  width=\columnwidth,
  height=4.5cm,
  ylabel={Count},
  xlabel={Version},
  symbolic x coords={v1.0,v2.0,v3.0,v3.4,v4.0},
  xtick=data,
  ymin=0, ymax=12,
  legend style={at={(0.5,1.02)}, anchor=south, legend columns=3, font=\scriptsize},
  nodes near coords,
  nodes near coords style={font=\tiny},
  enlarge x limits=0.15,
  every axis plot/.append style={fill opacity=0.7},
]
\addplot[fill=blue!60] coordinates {(v1.0,2) (v2.0,2) (v3.0,3) (v3.4,4) (v4.0,4)};
\addplot[fill=red!60]  coordinates {(v1.0,2) (v2.0,3) (v3.0,5) (v3.4,5) (v4.0,5)};
\addplot[fill=green!60] coordinates {(v1.0,1) (v2.0,1) (v3.0,1) (v3.4,2) (v4.0,2)};
\legend{Docker containers, DynamoDB tables, Services}
\end{axis}
\end{tikzpicture}
\caption{Infrastructure component growth across major versions. The jump at v3.0 reflects the Redis addition for BullMQ; v3.4 reflects the monolith-to-microservice decomposition; v4.0 adds the Operational API container.}
\label{fig:complexity}
\end{figure}

Figure~\ref{fig:complexity} tracks the growth of infrastructure components across versions.

\subsection{YouTube API Dependency Reduction}

\begin{figure}[h]
\centering
\begin{tikzpicture}
\begin{axis}[
  width=\columnwidth,
  height=4cm,
  ylabel={API calls/day (estimated)},
  xlabel={Version},
  symbolic x coords={v1.0,v1.4,v3.2,v3.3,v4.0},
  xtick=data,
  ymin=0, ymax=11000,
  mark size=3pt,
  nodes near coords,
  nodes near coords style={font=\tiny, above},
  every axis plot/.append style={thick},
]
\addplot[color=red, mark=*] coordinates {
  (v1.0,10000) (v1.4,9500) (v3.2,5000) (v3.3,1000) (v4.0,0)
};
\end{axis}
\end{tikzpicture}
\caption{Estimated daily YouTube API call reduction across versions. By v4.0, API consumption dropped to zero through the self-hosted Operational API.}
\label{fig:api-reduction}
\end{figure}
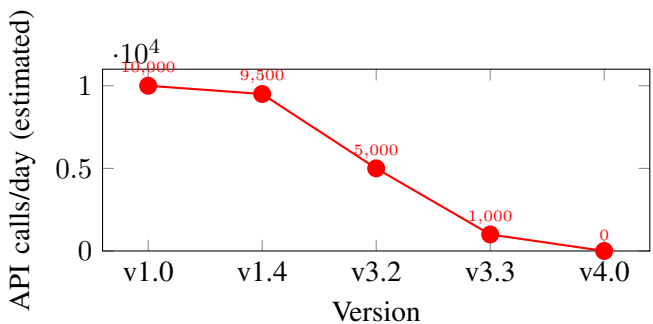

\subsection{Version Release Timeline}

Table~\ref{tab:versions} provides the complete version history with the architectural change and the production requirement that motivated each release.

\begin{table*}[t]
\centering
\caption{Complete Version History with Motivating Production Requirements}
\label{tab:versions}
\footnotesize
\begin{adjustbox}{max width=\textwidth}
\begin{tabular}{@{}llll@{}}
\toprule
\textbf{Version} & \textbf{Date} & \textbf{Key Changes} & \textbf{Forcing Function} \\
\midrule
1.0.0  & Aug 2022  & Initial release, full YouTube API dependency                     & MVP delivery requirement \\
1.1.0  & Late 2022 & AsyncLock on all DB ops, separate polling/processing intervals   & DynamoDB race condition bugs in production \\
1.2.0  & Late 2022 & PAY\_PER\_REQUEST billing, publish-date ordering                 & Throughput exceeded $\rightarrow$ 28 duplicate videos \\
1.4.1  & 2023      & Age-restricted exclusion, \texttt{maxRedirects:0} for uploads    & OOM errors from buffering 18\,GB payloads \\
1.5.0  & 2023      & Faucet integration, disk space check                             & Service crashes from full disk \\
2.0.0  & 2023      & Per-channel sync limits, verified-only syncing                   & Storage cost scaling at 10K+ channels \\
3.0.0  & Oct 2023  & BullMQ migration, YPP 2.0 tiers, batch TX                       & \texttt{better-queue} bottleneck \\
3.2.0  & Oct 2023  & yt-dlp for video metadata                                        & API quota exhaustion \\
3.3.0  & Nov 2023  & Chisel proxy, yt-dlp for channel info                            & YouTube IP blocking \\
3.4.0  & Feb 2024  & OpenTelemetry, service split                                     & Observability gap + scaling needs \\
3.6.0  & 2024      & Orion replaces Query Node                                        & Deprecated WebSocketLink \\
3.8.0  & 2024      & Yarn $\rightarrow$ npm, Polkadot.js Nara update                  & Tooling \& runtime compatibility \\
4.0.0  & 2024      & API-free signup, Operational API                                 & OAuth mass-expiration (10K+ channels) \\
\bottomrule
\end{tabular}
\end{adjustbox}
\end{table*}

\section{Multi-Operator Considerations}
\label{sec:multi-operator}

An important architectural question explored during the system's design was whether the YPP should support a single operator or multiple independent operators.

\textbf{Single-operator benefits}: No duplicate channel/video creation, consistent category mapping, simpler resource allocation, easier data migration and bug fixes.

\textbf{Multi-operator benefits}: Censorship resistance, no single point of failure, permissionless participation, no inter-operator coordination overhead.

The current design assumes a single operator, with the rationale that early program operators were all in organizational proximity within the Joystream DAO. Several mechanisms were proposed for future multi-operator support:

\begin{enumerate}[nosep]
  \item \textbf{Convention-based signaling}: Collaborator member naming conventions (e.g., \texttt{yt-synch::<ytChannelId>}) to signal ongoing synchronization
  \item \textbf{Checksum deduplication}: Content hash comparison across operators to prevent duplicate objects
  \item \textbf{Trusted list}: Dedicated registry of authorized gateway operators
  \item \textbf{Economic incentives}: Gateway tokens with slashing conditions for dishonest behavior
\end{enumerate}

The cooperative synching problem---where multiple operators must coordinate without a central authority while avoiding both duplication and starvation---remains an open research challenge.

\section{Discussion}
\label{sec:discussion}

\subsection{Lessons Learned}

\textbf{Defense layers are coupled, not independent}. The most damaging incidents arose not from a single YouTube defense mechanism, but from unforeseen interactions between them. The API quota forced a migration to \texttt{yt-dlp}, which made OAuth tokens dormant, which triggered Google's 6-month expiration policy, which opted out 10{,}000+ channels. Treating platform defenses as independent obstacles is architecturally dangerous; each circumvention must be evaluated for its side effects on other defense layers.

\textbf{Survivability beats throughput}. The 25$\times$ download concurrency reduction (50 $\rightarrow$ 2) with 0--90\,s random sleep per video appears to drastically limit throughput. In practice, a system that downloads at 2\% of peak speed but operates 100\% of the time extracts more content than one that runs at full speed for hours before being blocked for days. The operational cost of IP blocking---reprovisioning proxies, waiting out exclusion periods, manual intervention---far exceeds the cost of slower steady-state operation.

\textbf{Test infrastructure must match production}. The DynamoDB throughput incident (28 duplicate videos) was caused by testing against local DynamoDB without capacity limits. Production infrastructure characteristics---rate limits, throughput caps, latency profiles---must be reflected in test environments.

\textbf{Token lifecycle management requires active monitoring}. The 10K+ channel opt-out was caused by tokens silently expiring during a period when they were not being used. Systems that store third-party authentication tokens must either actively refresh them or decouple entirely from token-based authentication. The safest strategy, as YouTube-Synch ultimately demonstrated, is complete elimination.

\textbf{Schema changes propagate deeply}. The v4.0 schema cleanup (removing 11 OAuth-related fields across 2 tables) required updating the Pulumi IaC templates, API controllers, repository layer, startup reconciliation logic, Docker compose, and configuration schema. Coupling between schema design and system architecture is multiplicative.

\subsection{Limitations}

\begin{itemize}[nosep]
  \item As a retrospective experience report on a production system whose telemetry is not publicly releasable, we rely on operational incidents and first-principles analysis rather than controlled benchmarks; quantitative measurement of the detection arms race is left to future work.
  \item The self-hosted YouTube Operational API introduces a new external dependency whose long-term availability is uncertain; YouTube may patch the access patterns it relies on.
  \item The anti-detection parameters (sleep intervals, concurrency limits) were tuned empirically through production observation rather than through formal modeling of YouTube's detection algorithms, which are opaque.
  \item The livestream vs.\ premiere video discrimination problem remains unsolved, causing an unknown quantity of premiere videos to be incorrectly excluded.
  \item The priority scheduling algorithm's weights (Section~\ref{sec:priority}) were chosen empirically rather than through formal optimization.
  \item The single-operator assumption limits censorship resistance; multi-operator coordination remains an unsolved protocol design problem.
  \item Legal and ethical dimensions of creator-authorized bypass of YouTube's Terms of Service are not addressed in this paper.
\end{itemize}

\section{Conclusion and Future Work}
\label{sec:conclusion}

We presented YouTube-Synch, a production system that demonstrates the feasibility of automated, large-scale content extraction from a multi-billion dollar centralized video platform to decentralized blockchain infrastructure. Over 3.5 years and 15 release versions, the system evolved from a prototype fully dependent on YouTube's official API to an architecture that achieves \emph{zero} YouTube API consumption, \emph{zero} Google OAuth dependencies, and continuous operation through a pool-based proxy infrastructure with behavioral variance injection.

The central finding is that YouTube's content protection infrastructure, while formidable in isolation, can be systematically circumvented through sustained architectural adaptation---but only if the system treats each defense layer as part of an interconnected defense-in-depth system rather than independent obstacles. The three critical production incidents analyzed in this paper each resulted from unexpected interactions \emph{between} YouTube's defense layers: the API migration that rendered OAuth tokens unused (activating Google's token expiration policy), the DynamoDB billing configuration that interacted with blockchain immutability (producing irrecoverable duplicate objects), and the proxy infrastructure that required behavioral variance to avoid triggering a detection system calibrated for the traffic patterns of the previous proxy generation.

Key findings include: (1) complete decoupling from the source platform's official API is both achievable and beneficial---the transition from 10{,}000 daily API units to zero consumption eliminated an entire class of operational constraints; (2) anti-detection at scale requires sacrificing throughput for survivability---the 25$\times$ reduction in download concurrency (50 $\rightarrow$ 2) with triple random sleep injection demonstrates that slow, human-like access patterns are more sustainable than optimized parallel downloading; (3) authentication dependencies with centralized platforms create hidden coupling through token lifecycle policies, making full OAuth elimination the only architecturally safe long-term strategy; (4) fault tolerance in cross-platform replication requires treating the decentralized target (blockchain) and the centralized source (YouTube) as independently-failing systems that must be reconciled on every restart through a Write-Ahead Log pattern with cross-system state verification.

\textbf{Future work directions}:
\begin{itemize}[nosep]
  \item \textbf{Verifiable attribution}: Implementation of zkSNARK-based TLS oracle proofs (building on DECO~\cite{deco} and TLS-N~\cite{tlsn}) for non-interactive channel ownership verification that does not require trusting the YPP operator
  \item \textbf{Multi-operator coordination}: Formal protocol for cooperative synching without a central authority, including convention-based signaling, content hash deduplication, and economic incentives with slashing conditions
  \item \textbf{Livestream discrimination}: Reliable pre-ingestion filtering to distinguish active livestreams from premiere videos and recorded content
  \item \textbf{Multi-platform support}: Extension of the circumvention framework to other centralized platforms (Twitch, TikTok) with different defense layer configurations
  \item \textbf{Adaptive anti-detection}: Machine learning-based traffic shaping that responds to detection signal changes rather than relying on static behavioral variance parameters
  \item \textbf{Subtitle synchronization}: Automated transfer of closed captions and subtitle tracks across platforms
\end{itemize}


\balance

\end{document}